\documentclass[conference]{IEEEtran}
\ifCLASSINFOpdf
\else
\fi
\usepackage{graphicx}
\usepackage{setspace}
\usepackage{amsmath}
\usepackage{amssymb} 
\usepackage{bm} 
\usepackage{cite}
\usepackage{float}
\usepackage[usenames,dvipsnames]{pstricks}
\usepackage{epsfig}
\usepackage{pst-grad} 
\usepackage{pst-plot} 
\usepackage{array}
\newcommand{\lb} {\left}
\newcommand{\rb} {\right}
\newcommand{\nn} {\nonumber}
\specialpapernotice{(Invited Paper)}
\begin{document}
\title{Secrecy Performance of Dual-hop Threshold Relaying System with Diversity Reception}

\author{Chinmoy Kundu, Telex M. N. Ngatched, and Octavia A. Dobre \\Faculty of Engineering and Applied Science
\\Memorial University, St. John's, NL A1B 3X5, Canada \\ckundu@mun.ca,
tngatched@grenfell.mun.ca, and odobre@mun.ca}


\maketitle

\begin{abstract}
In this paper, the secrecy of a cooperative system consisting of a single source, relay, destination and eavesdropper is analyzed.
The threshold-selection decode-and-forward relay is considered, where the relay can correctly decode and forward only if it
satisfies a threshold signal-to-noise ratio (SNR). Both destination and eavesdropper take advantage of the direct and relayed
transmissions through maximal ratio diversity combining. The secrecy outage probability (SOP) and ergodic secrecy rate are
derived in closed-form for different channel state information (CSI) availability. It was observed that when the required rate is
low, having CSI knowledge is more advantageous than in the case of higher rate.  An increase in the required threshold SNR at the
relay can increase the SOP if the relayed link SNR is relatively higher than the direct link SNR. It was also shown that SOP
cannot be improved beyond a certain value when keeping either dual-hop link average SNR fixed and increasing the other link SNR,
whereas the ergodic secrecy rate can be increased by keeping the source to destination average SNR fixed.
\end{abstract}

\begin{IEEEkeywords}
Decode-and-forward relay, diversity technique, ergodic secrecy rate, maximal ratio combining, secrecy outage, threshold relaying.
\end{IEEEkeywords}
\section{Introduction}
\label{sec_intro}
The wireless medium is broadcast in nature, and hence, it is vulnerable to unintended eavesdropping.
Due to this reason, researchers are more and more interested
to study physical layer techniques to achieve information theoretic security by avoiding
upper layer data encryption \cite{wyner_wiretap, poor_infor_theo_sec,McLaughlin_wireless_info_theo_sec}.

With the advancement of cooperative relaying in wireless communication systems,
such schemes have also become popular to improve security
\cite{Petropulu_Poor_Impr_Wire_Phylay_Sec, Petropulu_On_Coop_Rel_Scheme}.
Cooperation is generally established via amplify-and-forward (AF) or decode-and-forward (DF) relays \cite{Laneman_Wornell_cooperative_diversity}.

DF relays can resist noise propagation to the subsequent stage, and hence, they have gained more importance in physical layer security over AF relays in dual-hop cooperative systems
\cite{krikidis_iet_opport_rel_sel, krikidis_twc_Rel_Sel_Jam, Bao_Relay_Selection_Schemes_Dual_Hop_Security, Alotaibi_Relay_Selection_MultiDestination,
Poor_Security_Enhancement_Cooperative, Hui_Secure_Relay_Jammer_Selection, Zou_Wang_Shen_optimal_relay_sel, Kundu_relsel, sarbani_Kundu_threshold_relay, Qahtani_relsel}.
Authors in \cite{krikidis_iet_opport_rel_sel, krikidis_twc_Rel_Sel_Jam, Bao_Relay_Selection_Schemes_Dual_Hop_Security, Alotaibi_Relay_Selection_MultiDestination,
Poor_Security_Enhancement_Cooperative, Hui_Secure_Relay_Jammer_Selection, Zou_Wang_Shen_optimal_relay_sel} consider
perfect decoding at the first hop, assuming high signal-to-noise ratio (SNR) for the source to relay link.
This assumption neglects the effect of fading associated loss of data rate in the first hop.
Deviating from this assumption, the authors in \cite{Kundu_relsel, sarbani_Kundu_threshold_relay, Qahtani_relsel}
 incorporate the first hop link quality into analysis. In \cite{Kundu_relsel}, the rate at the destination is evaluated as the minimum of the dual-hop.
 No direct links from the source to destination
 and eavesdropper are considered. The secrecy outage probability (SOP) of various relay selection schemes is analyzed.
 In \cite{sarbani_Kundu_threshold_relay}, the threshold-selection DF relay which is proposed in \cite{liu_opt_threshold},
 is used for the analysis of the SOP and ergodic secrecy rate. For the threshold-selection DF relay,
 perfect decoding is possible only if the instantaneous SNR exceeds a threshold at the relay. The authors do not consider the
 direct link from source to destination. In \cite{Qahtani_relsel}, perfect decoding is assumed at the relay only if the first hop can support a minimum threshold data rate.
 Authors in this paper study the secrecy enhancement through relay selection.
  Though \cite{Qahtani_relsel} takes into account both direct links from source to destination and eavesdropper, the authors do not explicitly investigate the threshold-selection DF relay.
 Furthermore, they concentrate on the performance of relay selection issues, and not on the performance of single relay system.
 The SOP when channel state information (CSI) knowledge is not available and the ergodic secrecy rate when CSI is available are studied.

 In this paper, we analyze a cooperative communication system with a single source, relay, destination, as well as a passive eavesdropper.
For the threshold-selection DF relay, we obtain closed-form expression of the SOP and ergodic
secrecy rate while having the direct links from the source to destination and the source to eavesdropper.
We evaluate the SOP and ergodic secrecy rate performances when CSI knowledge is available, as well as when it is not.
Additionally, we derive asymptotic performances when dual-hop links have unequal SNR and the required SNR threshold at the relay
asymptotically increases.

The remainder of this paper is organized as follows. Section \ref{sec_system} describes
the system model. Section \ref{sec_sop} presents the closed-form expressions of SOP for both known and unknown CSI.
In Section \ref{sec_erg_sec_rate}, the ergodic secrecy rate is derived. The asymptotic SOP is analyzed in Section \ref{sec_asymp_ana},
numerical and simulation results are provided in Section \ref{sec_num_result}, and conclusions are drawn in Section \ref{sec_conclusion}.

\textit{Notation:} $\mathbb{P}[\cdot]$ is the probability of occurrence
of an event, $\mathbb{E}_X[\cdot]$ defines the expectation of its argument over the random variable (R.V.)
$X$, $(x)^+\triangleq \max(0,x)$ and $\max{(\cdot)}$ denotes the maximum
of its argument, $F_X (\cdot)$ represents the
cumulative distribution function (CDF) of a R.V. $X$, and $f_X (\cdot)$ is the corresponding probability density function (PDF).

\begin{figure} []
\centering
\includegraphics[width=0.3\textwidth] {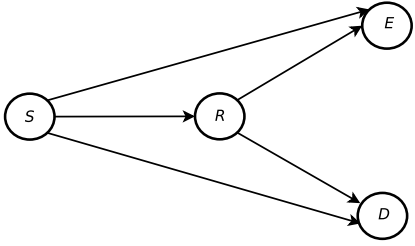}
\vspace*{-0.1cm}
\caption{System model with threshold-selection DF relay.}
\label{FIG_1}
\vspace{-.3cm}
\end{figure}

\section{System Model}
\label{sec_system}

The system model  consists of one source ($S$), one destination ($D$), one DF relay ($R$)
and one passive eavesdropper ($E$), as depicted in Fig. \ref{FIG_1}. $R$ is a threshold-selection DF relay of half-duplex transmission capability \cite{liu_opt_threshold}.
$S$ broadcasts its message in the first time slot. $D$, as well as $E$, receive the
direct broadcast from $S$. In the second time slot, if the SNR at $R$ is greater than the threshold SNR, $\gamma_{th}$,
it transmits the information. The SNR threshold can be set to ensure perfect decoding at the relay \cite{liu_opt_threshold}. $R$ does not transmit the received
message if it cannot decode the message correctly. To take the best advantage of two independent diverse paths, both
destination and eavesdropper consider  maximal ratio combining (MRC). The channels are modeled as independent flat Rayleigh fading.
The received SNR, $\gamma_{xy}$, of any arbitrary
$x$-$y$ link from node $x$ to node $y$ can be expressed as
$\gamma_{xy} = {P_{x}|h_{xy}|^2 }/{N_{0_{y}}}$,
where $x$ and $y$ are from $\{s, r, d, e\}$ corresponding to $\{S, R, D, E\}$ for any possible combination of $x$-$y$, $x\ne y$. $P_{x}$ is
the transmit power from node $x$ and $N_{0_{y}}$ is the noise variance of the
additive white Gaussian noise (AWGN) at node $y$. As $h_{xy}$ is assumed Rayleigh distributed with
average power unity, i.e., $\mathbb{E}[|h_{xy}|^2]=1$, $\gamma_{xy}$ is exponentially distributed
with mean $1/\lambda_{xy}=P_{x}/N_{0_{y}}$. The CDF of $\gamma_{xy}$ can be written as
\begin{align}
\label{eq_2}
F_{\gamma_{xy}}(z)=1-\exp(-\lambda_{xy}z), z\geq 0,
\end{align}
where $\lambda_{xy}$ is the parameter of exponential distribution. For notational simplicity, we further use the notation
$\lambda_{se}=\alpha_{se}$, $\lambda_{re}=\alpha_{re}$, $\lambda_{sr}=\beta_{sr}$, and $\lambda_{rd}=\beta_{rd}$.

The achievable secrecy rate is then given by \cite{wyner_wiretap}
\begin{align}
\label{eq_3}
C_s\triangleq{\frac{1}{2}\lb[\log_2\lb(\frac{1+\gamma_{M}}{1+\gamma_{E}}\rb)\rb]}^+,
\end{align}
where $\gamma_{M}$,  $\gamma_{E}$ are the SNR at $D$ and $E$, respectively.
The term $1/2$ shows that two time slots are necessary for information transfer.

The SOP is defined as the probability that the instantaneous
secrecy capacity is less than a target secrecy rate $R_s > 0$, i.e.,
\begin{align}
\label{eq_4}
P_o\lb(R_s\rb)&=\mathbb{P}\left[C_s < R_s  \right] = \mathbb{P}\lb[\gamma_{M} < \rho\lb(1+\gamma_{E}\rb)-1 \rb] \nn \\
&=\mathbb{E}_{\gamma_E}\lb[F_{\gamma_M}\lb(\rho\lb(1+\gamma_E\rb)-1\rb)\rb],
\end{align}
where $\rho = 2^{2R_s}$.

\section{Secrecy Outage Probability}
\label{sec_sop}
SOP is derived for two scenarios: when CSI is available and when CSI is not available
at the transmitters, $S$ and $R$. Henceforth, $S$ and $R$ are referred to as transmitters.
When $\gamma_{sr}>\gamma_{th}$, $R$ correctly decodes the message and as a result, $R$-$E$
and $R$-$D$ links exist. When $\gamma_{sr}<\gamma_{th}$, the $R$ does not transmit,
and only $S$-$E$ and $S$-$D$ link exist. As MRC diversity combining is performed at $D$ and $E$, when $\gamma_{sr}>\gamma_{th}$
we can write
\begin{align}
\label{eq_16}
\gamma_{M}&=\gamma_{sd}+\gamma_{rd},~~\gamma_{E}=\gamma_{se}+\gamma_{re}.
\end{align}
On the other hand, when $\gamma_{sr}<\gamma_{th}$, $\gamma_M$ and $\gamma_E$ are given as
\begin{align}
\label{eq_10}
\gamma_{M}&=\gamma_{sd},~~\gamma_{E}=\gamma_{se}.
\end{align}
The distribution of $\gamma_{M}$ and $\gamma_{E}$ can be obtained following the summation of two arbitrary independent exponentially
distributed random variables with different parameters as \cite{sum_expo_mohamed_akkouchi}
\begin{align}
\label{eq_5}
f(x)=\frac{\lambda_1\lambda_2}{\lambda_1-\lambda_2} e^{-\lambda_2 x}
+\frac{\lambda_1\lambda_2}{\lambda_2-\lambda_1} e^{-\lambda_1 x} ,
\end{align}
where $\lambda_1, \lambda_2$ are the parameters of the exponential distributions
of $X_1$ and $X_2$, respectively, with $\lambda_1 \neq \lambda_2$.
The corresponding CDF can be obtained by integrating (\ref{eq_5}) as
\begin{align}
\label{eq_6}
F(x)=1-\frac{\lambda_1}{\lambda_1-\lambda_2}e^{-\lambda_2 x}
-\frac{\lambda_2}{\lambda_2-\lambda_1}e^{-\lambda_1 x}.
\end{align}

\subsection{No CSI Knowledge}
\label{subsec_unknown_csi_outage}
When no CSI knowledge is available at the transmitters, these cannot adapt their rate
according to the CSI. Hence, the derivation of SOP is independent of the CSI.
The SOP of the system can be evaluated by finding the conditional SOP when the relay correctly
decodes the message and when it does not. From the theory of total probability, SOP
can be obtained as
\begin{align}
\label{eq_12}
&P_o(R_s) =\mathbb{P}\lb[C_s<R_s|\gamma_{sr}>\gamma_{th}\rb]\mathbb{P}\lb[\gamma_{sr}>\gamma_{th}\rb] \nn\\
&+\mathbb{P}\lb[C_s<R_s|\gamma_{sr}<\gamma_{th}\rb]\mathbb{P}\lb[\gamma_{sr}<\gamma_{th}\rb]\nn \\
&=\int_o^\infty F_{\gamma_M}\lb(\rho\lb(1+x\rb)-1\rb|\gamma_{sr}>\gamma_{th})f_{\gamma_E}(x|\gamma_{sr}>\gamma_{th})dx  \nn \\
&\times\lb[1-F_{\gamma_{sr}}\lb(\gamma_{th}\rb)\rb] +
\int_o^\infty F_{\gamma_M}\lb(\rho\lb(1+x\rb)-1\rb|\gamma_{sr}<\gamma_{th}) \nn\\
& \times f_{\gamma_E}(x|\gamma_{sr}<\gamma_{th})dx F_{\gamma_{sr}}\lb(\gamma_{th}\rb).
\end{align}
The above equation can be solved using (\ref{eq_16}) and (\ref{eq_10}) and their corresponding CDF and PDF.
The solution is expressed in (\ref{eq_20}), provided in the last page of the paper.

\subsection{Complete CSI Knowledge}
\label{known_csi_outage}
When CSI knowledge of all links is available at the transmitters, these can adapt their
rate according to the channel conditions to achieve positive secrecy.
From the theorem of total probability, SOP can be evaluated similar to the previous section. We can obtain SOP by imposing the
condition of positive secrecy, i.e., $\gamma_M>\gamma_E$, as
\begin{align}
\label{eq_23}
&P_o(R_s)=\mathbb{P}\lb[C_s<R_s\cap\gamma_M>\gamma_E|\gamma_{sr}>\gamma_{th}\rb]
\mathbb{P}\lb[\gamma_{sr}>\gamma_{th}\rb] \nn\\
&+\mathbb{P}\lb[C_s<R_s\cap\gamma_M>\gamma_E|\gamma_{sr}<\gamma_{th}\rb]
\mathbb{P}\lb[\gamma_{sr}<\gamma_{th}\rb].
\end{align}
The quantity $\mathbb{P}\lb[C_s<R_s\cap\gamma_M>\gamma_E|\gamma_{sr}>\gamma_{th}\rb]$
is evaluated as
\begin{align}
\label{eq_24}
&\mathbb{P}\lb[C_s<R_s\cap\gamma_M>\gamma_E|\gamma_{sr}>\gamma_{th}\rb]\nn \\
&=\mathbb{P}\lb[\gamma_E<\gamma_M<\rho(1+\gamma_E)-1|\gamma_{sr}>\gamma_{th}\rb] \nn \\
&=\int_0^\infty\int_y^{\rho(1+y)-1} f_{\gamma_M}(x) f_{\gamma_E}(y)dx dy \nn \\
&=\int_0^\infty[F_{\gamma_M}\lb(\rho\lb(1+y\rb)-1\rb)-F_{\gamma_M}(y)]f_{\gamma_E}(y) dy .
\end{align}
The final solution is given in (\ref{eq_26}), placed in the last page of the paper.
\section{Ergodic Secrecy Rate}
\label{sec_erg_sec_rate}
In this section, the ergodic secrecy rate of the system is evaluated with the assumption of complete CSI knowledge and no CSI knowledge
at the transmitters, respectively. The ergodic secrecy rate of the system, $\bar{C}_s$, can be expressed as \cite{Olabiyi_Sec}
\begin{align}
\label{eq_29}
\bar{C}_s
&=\bar{C}_s({\gamma_{sr}\geq\gamma_{th}})\mathbb{P}\lb[\gamma_{sr}\geq\gamma_{th}\rb] \nn \\
&+\bar{C}_s({\gamma_{sr}<\gamma_{th}})\mathbb{P}\lb[\gamma_{sr}<\gamma_{th}\rb] \nn \\
&=\bar{C}_s({\gamma_{sr}\geq\gamma_{th}})\lb(1-\mathbb{P}\lb[\gamma_{sr}<\gamma_{th}\rb]\rb)\nn \\
&+\bar{C}_s({\gamma_{sr}<\gamma_{th}})\mathbb{P}\lb[\gamma_{sr}<\gamma_{th}\rb],
\end{align}
where $\bar{C}_s({\gamma_{sr}\geq\gamma_{th}})$ is the conditional ergodic secrecy rate when
$\gamma_{sr}\geq\gamma_{th}$, and similarly, $\bar{C}_s({\gamma_{sr}<\gamma_{th}})$ is the
conditional ergodic secrecy rate when $\gamma_{sr}<\gamma_{th}$.
$\mathbb{P}\lb[\gamma_{sr}<\gamma_{th}\rb]$ can be evaluated from (\ref{eq_2}).
\subsection{No CSI Knowledge}
\label{unknown_csi}
$\bar{C}_s({\gamma_{sr}\geq\gamma_{th}})$ can be evaluated from (\ref{eq_3}) as
\begin{align}
\label{eq_30}
&\bar{C}_s({\gamma_{sr}\geq\gamma_{th}})
=\frac{1}{2\ln2}\int_0^\infty \int_0^\infty{\ln\lb[\frac{1+x}{1+y}\rb]}
f_{\gamma_{M}}(x|_{\gamma_{sr}\geq\gamma_{th}}) \nn \\
&\times f_{\gamma_{E}}(y|_{\gamma_{sr}\geq\gamma_{th}}) dxdy\nn \\
&=\frac{1}{2\ln2} \lb[\bar{I}_M({\gamma_{sr}\geq\gamma_{th}})-\bar{I}_E({\gamma_{sr}\geq\gamma_{th}})\rb],
\end{align}
where $\bar{I}_M({\gamma_{sr}\geq\gamma_{th}})$, $\bar{I}_E({\gamma_{sr}\geq\gamma_{th}})$ are expressed as
\begin{align}
\label{eq_31}
&\bar{I}_M(\gamma_{sr}\geq\gamma_{th})
=\int_0^\infty \ln\lb(1+x\rb)f_{\gamma_M}(x\mid_{\gamma_{sr}\geq\gamma_{th}})dx, \\
\label{eq_32}
&\bar{I}_E(\gamma_{sr}\geq\gamma_{th})
=\int_0^\infty \ln\lb(1+y\rb)f_{\gamma_E}(y|{\gamma_{sr}\geq\gamma_{th}})dy,
\end{align} respectively.
Now, $\bar{C}_s\lb({\gamma_{sr}<\gamma_{th}}\rb)$ can be evaluated in a similar way as
(\ref{eq_30})-(\ref{eq_32}). The final expression is given in ($\ref{eq_38}$), written in the last page of the paper.
\subsection{Complete CSI Knowledge}
\label{known_csi}
From (\ref{eq_3}), $\bar{C}_s({\gamma_{sr}\geq\gamma_{th}})$ can be evaluated as
\begin{align}
\label{eq_33}
&\bar{C}_s({\gamma_{sr}\geq\gamma_{th}})
=\frac{1}{2\ln2}\int_0^\infty \int_0^x {\ln\lb[\frac{1+x}{1+y}\rb]} f_{\gamma_{M}}(x|{\gamma_{sr}\geq\gamma_{th}})\nn\\
&\times f_{\gamma_{E}}(y|{\gamma_{sr}\geq\gamma_{th}}) dxdy\nn \\
&=\frac{1}{2\ln2}\lb[\bar{I}_M({\gamma_{sr}\geq\gamma_{th}})-\bar{I}_E({\gamma_{sr}\geq\gamma_{th}})\rb],
\end{align}
where $\bar{I}_M({\gamma_{sr}\geq\gamma_{th}})$ is $\bar{I}_M$ conditioned on $\gamma_{sr}\geq\gamma_{th}$
and $\bar{I}_E({\gamma_{sr}\geq\gamma_{th}})$  is $\bar{I}_E$ conditioned on $\gamma_{sr}\geq\gamma_{th}$.
$\bar{I}_M({\gamma_{sr}\geq\gamma_{th}})$ and $\bar{I}_E({\gamma_{sr}\geq\gamma_{th}})$ can be expressed as
\begin{align}
\label{eq_34}
\bar{I}_M(\gamma_{sr}\geq\gamma_{th})
&=\int_0^\infty \int_0^x \ln\lb(1+x\rb)f_{\gamma_E}(y|{\gamma_{sr}\geq\gamma_{th}}) \nn \\
& \times f_{\gamma_M}(x|{\gamma_{sr}\geq\gamma_{th}}) dydx, \\
\label{eq_35}
\bar{I}_E(\gamma_{sr}\geq\gamma_{th})
&=\int_0^\infty \int_0^x \ln\lb(1+y\rb)f_{\gamma_E}(y|{\gamma_{sr}\geq\gamma_{th}}) \nn \\
&\times f_{\gamma_M}(x|\gamma_{sr}\geq\gamma_{th}) dydx \nn \\
&=\int_0^\infty \ln(1+y)F_{\gamma_{E}}^c(y|\gamma_{sr}>\gamma_{th})\nn\\
&\times f_{\gamma_{E}}(y|\gamma_{sr}>\gamma_{th}) dy,
\end{align} respectively.
Now, $\bar{C}_s\lb({\gamma_{sr}<\gamma_{th}}\rb)$ can be evaluated in a similar way as (\ref{eq_33})-(\ref{eq_35}). The final expression is given
in (\ref{eq_39}), shown in the last page of the paper.

\section{Asymptotic Analysis}
\label{sec_asymp_ana}
\subsection{Asymptotic Thresholding}
\label{asymp_threshold}
It is important to find the SOP when $R$ is always able to decode the message, as well as when it is never able to do so.
If we allow $\gamma_{th} \rightarrow 0$, in this case, $R$ is able to decode the message correctly.
If we allow $\gamma_{th}\rightarrow \infty$,  $R$ is not able to decode the message correctly; hence, $R$-$D$ and $R$-$E$ links do not exist.
The asymptotic behavior of SOP when $\gamma_{th} \rightarrow 0$ can be evaluated by simply
replacing $\gamma_{th}= 0$ in (\ref{eq_20}) and (\ref{eq_26}). Similarly, when $\gamma_{th} \rightarrow \infty$, the asymptotic SOP can be
evaluated from (\ref{eq_20}) and (\ref{eq_26}) by replacing $\gamma_{th} = \infty$.
It is to be noted from (\ref{eq_20}) that when $\gamma_{th} \rightarrow \infty$, the SOP is the one of the wiretap channel
\cite{McLaughlin_wireless_info_theo_sec}.
\subsection{Balanced and Unbalanced Cases}
\label{asymp_balance}
The balanced case provides the behaviour of $P_o(R_s)$ when both dual-hop links are
very strong when compared with the direct links to $D$ and $E$. This can happen if $S$, $R$ and
$D$ are closely spaced compared to $E$. The condition is $1/\beta_{sr}=1/\beta_{rd}=1/\beta\rightarrow  \infty$.
The asymptotic expression is derived from (\ref{eq_20}) for the case when no CSI is known, and shown in (\ref{eq_48}).
The derivation is similar for the case of full CSI knowledge.

The unbalance case arises if either one of the $S-R$ or $R-D$ link is
closely spaced when compared with the other links.
When $1/\beta_{sr}$ is kept fixed and $1/\beta_{rd}=1/\beta\rightarrow  \infty$ or
$1/\beta_{rd}$ is kept fixed and $1/\beta_{sr}=1/\beta\rightarrow  \infty$,
the asymptotic SOP can be expressed as a summation of a constant quantity and an asymptotically
varying term.  At low SNR, the varying term dominates, while it vanishes at high SNR. The constant and asymptotic terms for both unbalanced cases are
derived for the case when CSI is unknown; the expressions are provided in (\ref{eq_51}) and (\ref{eq_55}), being derived from (\ref{eq_20}).

\section{Numerical and Simulation Results}
\label{sec_num_result}
This section discusses numerical and simulation results. The unit of required rate, $R_s$, is bits per channel use (bpcu). Without loss of generality,
we assume that all nodes are affected by the same noise power. The plots obtained when no CSI is available are marked as NOCSI; when CSI information is
available, they are marked as CSI. A general observation from all figures is that SOP is lower and the ergodic secrecy rate is higher when CSI is available
compared to the case when such information is not available. When CSI is available, the transmission rate can be changed accordingly to alleviate the
detrimental effect of fading.

\begin{figure}[]
\centering
\includegraphics[width=0.49\textwidth]{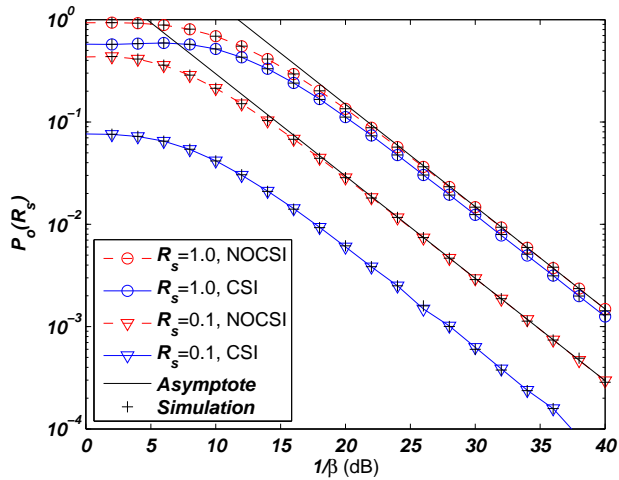}
\vspace*{-0.2cm}
\caption{Secrecy outage probability for different rate requirements.}
\label{FIG_VTC_2}
\vspace{0cm}
\end{figure}

In Fig. \ref{FIG_VTC_2}, SOP of the system is shown in the balanced case for two different rate requirements.
We assume that $1/\beta_{sr}=1/\beta_{rd}=1/\beta$. The results are obtained when
$\gamma_{th}=3$ dB, $1/\alpha_{se}=0$ dB, $1/\alpha_{re}=3$ dB and $1/\beta_{sd}=3$.
The asymptotes found in (\ref{eq_48}) are also plotted for the NOCSI case with solid line, tangent to the SOP plots. We can see that at lower required
rate, $R_s=0.1$, the performance gap between CSI and NOCSI is more than for the higher required rate, $R_s=1$. This shows that
for a lower required rate, CSI knowledge can improve the performance more when compared with the case when the rate requirement is high.

\begin{figure}[]
\centering
\includegraphics[width=0.49\textwidth]{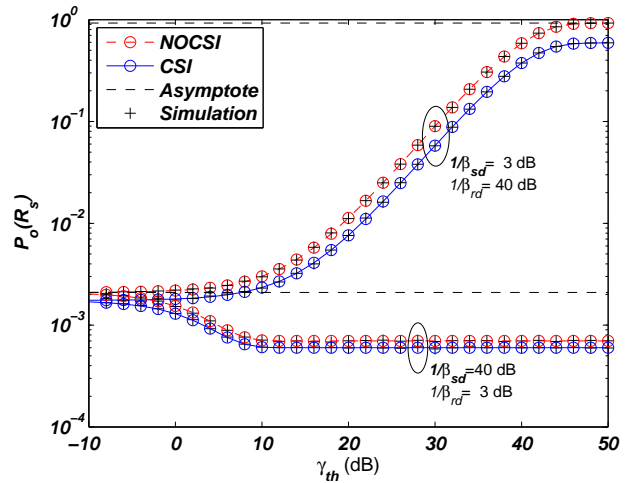}
\vspace*{-0.2cm}
\caption{Secrecy outage probability versus the SNR threshold at $R$.}
\label{FIG_VTC_4}
\vspace{-.3cm}
\end{figure}

In Fig. \ref{FIG_VTC_4}, SOP versus $\gamma_{th}$ is shown in the balanced case, at $R_s=1$. Two different conditions are of special importance:
i) when the direct link average SNR, $1/\beta_{sd}=40$ dB, is much higher than for the relayed link $1/\beta_{rd}=3$ dB;
ii) the relayed link SNR $1/\beta_{rd}=40$ dB is much higher than the direct link SNR $1/\beta_{sd}=3$ dB. The eavesdropper channel condition is kept the same for both
$1/\alpha_{se}=0$ dB and  $1/\alpha_{re}=6$ dB. The asymptotic values are also drawn with horizontal dashed lines.  It can be observed that if the relayed link
is better than the direct link, SOP increases with the increase in $\gamma_{th}$.  If we increase the required threshold at $R$, the probability that the relay can
exceed this threshold decreases. As a result, the probability that the relayed
transmission is blocked also increases. This is more disadvantageous for the system when the $R-D$ link SNR is much higher than the direct link SNR in case ii).
As a results, SOP increases. On the contrary, in case i), it is advantageous for the system from the security perspective if $R-E$ transmission is blocked due
to increase in $\gamma_{th}$. Hence SOP decreases in case i).

\begin{figure} []
\centering
\includegraphics[width=0.49\textwidth]{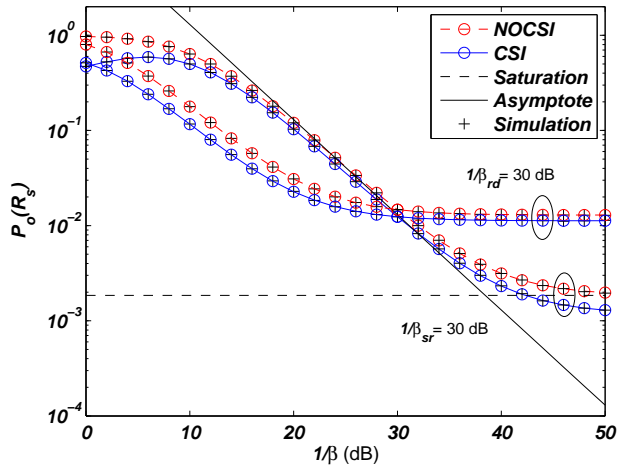}
\vspace*{-0.2cm}
\caption{Secrecy outage probability for unbalance in the main link.}
\label{FIG_VTC_5}
\vspace{-0cm}
\end{figure}

Fig. \ref{FIG_VTC_5} provides SOP results for the unbalanced case. Here, $R_s=1$, $\gamma_{th}=3$ dB,
$1/\alpha_{se}=0$ dB, $1/\alpha_{re}=3$ dB, and $1/\beta_{sd}=3$ dB. Both unbalance conditions:
i) keeping $1/\beta_{sr}=30$ dB fixed and increasing $1/\beta = 1/\beta_{rd}$ dB, ii) keeping $1/\beta_{rd}=30$ dB fixed and
increasing $1/\beta=1/\beta_{sr}$ dB are shown. We can see from the figure that SOP saturates to a particular value in both unbalanced cases.
The particular saturation values are shown by the horizontal dashed line. The asymptotic value through which it attains the saturation is shown by solid line.
Dashed lines are the constant terms in (\ref{eq_51}) and (\ref{eq_55}), whereas the solid line represents the term associated with $1/\beta$ in the same equation.
This observation leads to the conclusion that even if the SNR of $S-R$ or $R-D$ links increases, keeping other link SNR fixed cannot improve SOP.

\begin{figure}[]
\centering
\includegraphics[width=0.49\textwidth]{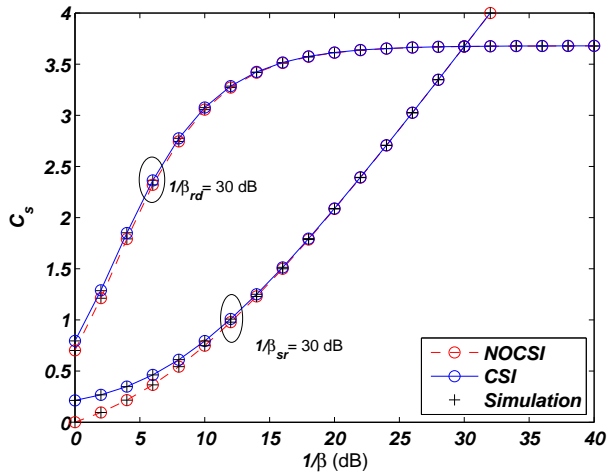}
\vspace*{-0.2cm}
\caption{Ergodic secrecy rate for unbalance in the main link.}
\label{FIG_VTC_6}
\vspace{-.3cm}
\end{figure}

Fig. \ref{FIG_VTC_6} depicts the ergodic secrecy rate for the unbalanced case. Here, the parameters are set-up as for results in Fig. \ref{FIG_VTC_5}.
It is observed that saturation does not occur in both unbalanced cases i) and ii), which is contrary to Fig. \ref{FIG_VTC_5}. Saturation occurs when
$1/\beta_{rd}$ is fixed. By increasing the $1/\beta_{sr}$, possibility of relayed communication increases; however it gives no advantage, as
$1/\beta_{rd}$ SNR is fixed. Hence, the ergodic secrecy rate saturates depending on the fixed value of $1/\beta_{rd}$. On the contrary, if $1/\beta_{rd}$ increases,
keeping $1/\beta_{sr}$ fixed, a higher rate still can be ensured through the $R-D$ channel.

\section{Conclusions}
\label{sec_conclusion}
The secrecy outage probability and ergodic secrecy rate of a threshold-selection DF relay system is evaluated
in closed-form for known and unknown CSI, respectively. Both eavesdropper and destination exploit the
diversity benefit from the direct transmission. The asymptotic analysis is provided when dual-hop links have equal or unequal
channel SNR. It is seen that known CSI is more beneficial when the required rate is low. It is also observed that if the required SNR threshold increases, the
secrecy outage probability can increase when the relayed link quality is better than the direct link quality. Additionally, it is showed that any of the dual-hop link quality
can be a bottleneck for the secrecy outage, while only the relay to destination channel quality can be a bottleneck for the ergodic secrecy rate.

\bibliographystyle{IEEEtran}
\bibliography{IEEEabrv,MYALL_REFERENCE}

\newcommand{\noop}[1]{}
\begin{thebibliography}{10}
\providecommand{\url}[1]{#1}
\csname url@samestyle\endcsname
\providecommand{\newblock}{\relax}
\providecommand{\bibinfo}[2]{#2}
\providecommand{\BIBentrySTDinterwordspacing}{\spaceskip=0pt\relax}
\providecommand{\BIBentryALTinterwordstretchfactor}{4}
\providecommand{\BIBentryALTinterwordspacing}{\spaceskip=\fontdimen2\font plus
\BIBentryALTinterwordstretchfactor\fontdimen3\font minus
  \fontdimen4\font\relax}
\providecommand{\BIBforeignlanguage}[2]{{%
\expandafter\ifx\csname l@#1\endcsname\relax
\typeout{** WARNING: IEEEtran.bst: No hyphenation pattern has been}%
\typeout{** loaded for the language `#1'. Using the pattern for}%
\typeout{** the default language instead.}%
\else
\language=\csname l@#1\endcsname
\fi
#2}}
\providecommand{\BIBdecl}{\relax}
\BIBdecl

\bibitem{wyner_wiretap}
A.~D. Wyner, ``{The Wire-Tap Channel},'' \emph{Bell System Technical Journal},
  vol.~54, no.~8, pp. 1355--1387, Oct. 1975.

\bibitem{poor_infor_theo_sec}
Y.~Liang, H.~V. Poor, and S.~Shamai, ``{Information Theoretic Security},''
  \emph{Foundations and Trends in Communications and Information Theory},
  vol.~5, no. 4--5, pp. 355--580, Jun. 2009.

\bibitem{McLaughlin_wireless_info_theo_sec}
M.~Bloch, J.~Barros, M.~R.~D. Rodrigues, and S.~McLaughlin, ``{Wireless
  Information-Theoretic Security},'' \emph{{IEEE} Trans. Inf. Theory}, vol.~54,
  no.~6, pp. 2515--2534, Jun. 2008.

\bibitem{Petropulu_Poor_Impr_Wire_Phylay_Sec}
L.~Dong, Z.~Han, A.~Petropulu, and H.~Poor, ``{Improving Wireless Physical
  Layer Security via Cooperating Relays},'' \emph{{IEEE} Trans. Signal
  Process.}, vol.~58, no.~3, pp. 1875--1888, Mar. 2010.

\bibitem{Petropulu_On_Coop_Rel_Scheme}
J.~Li, A.~Petropulu, and S.~Weber, ``{On Cooperative Relaying Schemes for
  Wireless Physical Layer Security},'' \emph{{IEEE} Trans. Signal Process.},
  vol.~59, no.~10, pp. 4985--4997, Oct. 2011.

\bibitem{Laneman_Wornell_cooperative_diversity}
{J. N. Laneman}, {D. N. C. Tse}, and {G. W. Wornell}, ``{Cooperative Diversity
  in Wireless Networks: Efficient Protocols and Outage Behavior},''
  \emph{{IEEE} Trans. Inf. Theory}, vol.~50, no.~12, pp. 3062--3080, Dec. 2004.

\bibitem{krikidis_iet_opport_rel_sel}
I.~Krikidis, ``{Opportunistic Relay Selection for Cooperative Networks with
  Secrecy Constraints},'' \emph{IET Commun.}, vol.~4, no.~15, pp. 1787--1791,
  Oct. 2010.

\bibitem{krikidis_twc_Rel_Sel_Jam}
I.~Krikidis, J.~Thompson, and S.~McLaughlin, ``{Relay Selection for Secure
  Cooperative Networks with Jamming},'' \emph{{IEEE} Trans. Wireless Commun.},
  vol.~8, no.~10, pp. 5003--5011, Oct. 2009.

\bibitem{Bao_Relay_Selection_Schemes_Dual_Hop_Security}
V.~N.~Q. Bao, {N. Linh-Trung}, and M.~Debbah, ``{Relay Selection Schemes for
  Dual-Hop Networks under Security Constraints with Multiple Eavesdroppers},''
  \emph{{IEEE} Trans. Wireless Commun.}, vol.~12, no.~12, pp. 6076--6085, Dec.
  2013.

\bibitem{Alotaibi_Relay_Selection_MultiDestination}
{E. R. Alotaibi} and K.~A. Hamdi, ``{Relay Selection for Multi-Destination in
  Cooperative Networks with Secrecy Constraints},'' in \emph{{Proc. IEEE
  Vehicular Technology Conference}}, Sep. 2014, pp. 1--5.

\bibitem{Poor_Security_Enhancement_Cooperative}
{L. Wang}, {K. J. Kim}, {T. Q. Duong}, {M. Elkashlan}, and H.~Poor, ``{Security
  Enhancement of Cooperative Single Carrier Systems},'' \emph{IEEE Trans. Inf.
  Forensics and Security}, vol.~10, no.~1, pp. 90--103, Jan. 2015.

\bibitem{Hui_Secure_Relay_Jammer_Selection}
{H. Hui}, {A. Swindlehurst}, {G. Li}, and J.~Liang, ``{Secure Relay and Jammer
  Selection for Physical Layer Security},'' \emph{IEEE Signal Process. Lett.},
  vol.~22, no.~8, pp. 1147--1151, Aug. 2015.

\bibitem{Zou_Wang_Shen_optimal_relay_sel}
Y.~Zou, X.~Wang, and W.~Shen, ``{Optimal Relay Selection for Physical-Layer
  Security in Cooperative Wireless Networks},'' \emph{{IEEE} J. Sel. Areas
  Commun.}, vol.~31, no.~10, pp. 2099--2111, Oct. 2013.

\bibitem{Kundu_relsel}
C.~Kundu, S.~Ghose, and R.~Bose, ``{Secrecy Outage of Dual-hop Regenerative
  Multi-Relay System with Relay Selection},'' \emph{{IEEE} Trans. Wireless
  Commun.}, vol.~14, no.~8, pp. 4614--4625, Aug. 2015.

\bibitem{sarbani_Kundu_threshold_relay}
\BIBentryALTinterwordspacing
S.~Ghose, C.~Kundu, and R.~Bose, ``{Secrecy Performance of Dual-hop DF Relay
  System with Diversity Combining at the Eavesdropper},'' \emph{IET
  Communications}, Feb. 2016. [Online]. Available:
  \url{http://dx.doi.org/10.1049/iet-com.2015.1060}
\BIBentrySTDinterwordspacing

\bibitem{Qahtani_relsel}
F.~S. AL-Qahtani, C.~Zhong, and H.~Alnuweiri, ``{Opportunistic Relay Selection
  for Secrecy Enhancement in Cooperative Networks},'' \emph{{IEEE} Trans.
  Commun.}, vol.~63, no.~5, pp. 1756--1770, May 2015.

\bibitem{liu_opt_threshold}
W.~Siriwongpairat, T.~Himsoon, W.~Su, and K.~Liu, ``{Optimum
  Threshold-Selection Relaying for Decode-and-Forward Cooperation Protocol},''
  in \emph{{Proc. IEEE Wireless Communications and Networking Conference}},
  Apr. 2006, pp. 1015--1020.

\bibitem{sum_expo_mohamed_akkouchi}
M.~Akkouchi, ``{On the Convolution of Exponential Distributions},'' \emph{J.
  Chungcheong Math. Soc.}, vol.~21, no.~4, pp. 501--510, Dec. 2008.

\bibitem{Olabiyi_Sec}
O.~Olabiyi and A.~Annamalai, ``{Ergodic Secrecy Rates of Secure Wireless
  Communications},'' in \emph{{Proc. IEEE Military Communications Conference
  (MILCOM)}}, Oct 2014, pp. 18--23.

\end{thebibliography}

\begin{table*}
In the following equations, $\operatorname{Ei(x)}=-\int_{-x}^{\infty}\exp(-t)/t dt$ \cite{Olabiyi_Sec}. \\\\
\label{table_sec}
\begin{tabular}{m{\textwidth}}
\hline\\
{\centering The SOP derived from (\ref{eq_12}) when no CSI knowledge is available.
\begin{align}
\label{eq_20}
P_o(R_s) &= 1-\frac{\alpha_{se}e^{-\beta_{sd}(\rho-1)}}{\alpha_{se}+\rho\beta_{sd}} +
e^{-(\beta_{sr}\gamma_{th})}\lb[\frac{\alpha_{se}e^{-\beta_{sd}(\rho-1)}}{\alpha_{se}+\rho\beta_{sd}}
- \frac{\beta_{sd}\alpha_{re}\alpha_{se}\exp{\lb(-\beta_{rd}(\rho-1)\rb)}}{(\beta_{sd}-\beta_{rd})
(\rho\beta_{rd}+\alpha_{re})(\rho\beta_{rd}+\alpha_{se})} \rb.\nn\\
&\lb.-\frac{\beta_{rd}\alpha_{re}\alpha_{se}\exp{\lb(-\beta_{sd}
(\rho-1)\rb)}}{(\beta_{rd}-\beta_{sd})(\rho\beta_{sd}+\alpha_{re})(\rho\beta_{sd}+\alpha_{se})} \rb].
\end{align}
}\\
{\centering The SOP derived from (\ref{eq_24}) when complete CSI knowledge is available.
\begin{align}
\label{eq_26}
P_o(R_s)
&=\frac{\alpha_{se}}{\alpha_{se}+\beta_{sd}}
-\frac{\alpha_{se}e^{-\beta_{sd}\lb(\rho-1\rb)}}{\alpha_{se}+\rho\beta_{sd}}
-e^{-(\beta_{sr}\gamma_{th})}
\lb[\frac{\alpha_{se}}{\alpha_{se}+\beta_{sd}}-\frac{\alpha_{se}e^{-\beta_{sd}(\rho-1)}}{\alpha_{se}+\rho\beta_{sd}}
+\frac{\beta_{sd}\alpha_{se}\alpha_{re}}
{\lb(\beta_{sd}-\beta_{rd}\rb)\lb(\alpha_{se}+\beta_{rd}\rb)\lb(\alpha_{re}+\beta_{rd}\rb)}\rb. \nn \\
&\lb.-\frac{\beta_{rd}\alpha_{se}\alpha_{re}}
{\lb(\beta_{rd}-\beta_{sd}\rb)\lb(\alpha_{se}+\beta_{sd}\rb)\lb(\alpha_{re}+\beta_{sd}\rb)}
+\frac{\beta_{sd}\alpha_{se}\alpha_{re}e^{-\beta_{rd}(\rho-1)}}
{\lb(\beta_{sd}-\beta_{rd}\rb)\lb(\alpha_{se}+\rho\beta_{rd}\rb)\lb(\alpha_{re}+\rho\beta_{rd}\rb)} \rb.\nn\\
&\lb.+\frac{\beta_{rd}\alpha_{se}\alpha_{re}e^{-\beta_{sd}(\rho-1)}}
{\lb(\beta_{rd}-\beta_{sd}\rb)\lb(\alpha_{se}+\rho\beta_{sd}\rb)\lb(\alpha_{re}+\rho\beta_{sd}\rb)}\rb].
\end{align}
}\\\hline\\
{
\centering The ergodic secrecy rate derived from (\ref{eq_30}) when no CSI knowledge is available.
\begin{align}
\label{eq_38}
\bar{C}_s
&=e^{\alpha_{se}}\operatorname{Ei}\lb(-\alpha_{se}\rb)-e^{\beta_{sd}}\operatorname{Ei}\lb(-\beta_{sd}\rb)
+e^{-(\beta_{sr}\gamma_{th})}
\lb[\frac{\beta_{sd}}{\beta_{sd}-\beta_{rd}}\lb(e^{\beta_{sd}}\operatorname{Ei}\lb(-\beta_{sd}\rb)
-e^{\beta_{rd}}\operatorname{Ei}\lb(-\beta_{rd}\rb)  \rb)\rb.\nn \\
&\lb.-\frac{\alpha_{se}}{\alpha_{se}-\alpha_{re}}\lb(e^{\alpha_{se}}\operatorname{Ei}\lb(-\alpha_{se}\rb)
-e^{\alpha_{re}}\operatorname{Ei}\lb(-\alpha_{re}\rb)\rb)\rb].
\end{align}
}
\\
{
\centering The ergodic secrecy rate derived from (\ref{eq_33}) when complete CSI knowledge is available.
\begin{align}
\label{eq_39}
\bar{C}_s
&=e^{\beta_{sd}+\alpha_{se}}\operatorname{Ei}\lb(-\beta_{sd}-\alpha_{se}\rb)
-e^{\beta_{sd}}\operatorname{Ei}\lb(-\beta_{sd}\rb)
+\frac{e^{-(\beta_{sr}\gamma_{th})}}{\lb(\beta_{sd}-\beta_{rd}\rb)}
\lb[\beta_{sd}\lb(e^{\beta_{sd}}\operatorname{Ei}\lb(-\beta_{sd}\rb)
-e^{\beta_{rd}}\operatorname{Ei}\lb(-\beta_{rd}\rb)\rb)\rb.\nn \\
&\lb.+\frac{1}{\lb(\alpha_{se}-\alpha_{re}\rb)}
\lb(\lb(\beta_{rd}\alpha_{se}+\beta_{sd}\alpha_{re}-\beta_{sd}\alpha_{se}\rb)
e^{\beta_{sd}+\alpha_{se}}\operatorname{Ei}\lb(-\beta_{sd}-\alpha_{se}\rb)
-\beta_{rd}\alpha_{se}e^{\beta_{sd}+\alpha_{re}}\operatorname{Ei}\lb(-\beta_{sd}-\alpha_{re}\rb) \rb.\rb.\nn \\
&\lb.\lb.-\beta_{sd}\alpha_{re}e^{\beta_{rd}+\alpha_{se}}\operatorname{Ei}\lb(-\beta_{rd}-\alpha_{se}\rb)
+\beta_{sd}\alpha_{se}e^{\beta_{rd}+\alpha_{re}}\operatorname{Ei}\lb(-\beta_{rd}-\alpha_{re}\rb) \rb)\rb].
\end{align}
}\\
\hline
\\
{\centering The asymptotic SOP derived from (\ref{eq_20}) in the balanced case.
\begin{align}
\label{eq_48}
P_o^{AS}(R_s)
&=\frac{1}{\frac{1}{\beta}}\lb[\gamma_{th}-1-\frac{1}{\beta_{sd}}
+\frac{\rho\lb(\alpha_{se}+\alpha_{re}+\alpha_{se}\alpha_{re}\rb)}{\alpha_{se}\alpha_{re}}
+\frac{\alpha_{se}e^{-\beta_{sd}\lb(\rho-1\rb)}}{\alpha_{se}+\rho\beta_{sd}}
\lb(\frac{\alpha_{re}}{\beta_{sd}\lb(\alpha_{re}+\rho\beta_{sd}\rb)}-\gamma_{th}\rb)\rb].
\end{align}
}\\\hline\\
{\centering The asymptotic SOP derived from  (\ref{eq_20}) when $1/\beta_{sr}$ is fixed and
$1/\beta_{rd}=1/\beta\rightarrow  \infty$.
\begin{align}
\label{eq_51}
P_o^{AS}(R_s)&
=\lb(1-e^{-\beta_{sr}\gamma_{th}}\rb)\lb(1-\frac{\alpha_{se}e^{-\beta_{sd}\lb(\rho-1\rb)}}{\alpha_{se}+\rho\beta_{sd}}\rb)
+\frac{1}{\frac{1}{\beta}}\lb[\frac{\rho\lb(\alpha_{se}+\alpha_{re}+\alpha_{se}\alpha_{re}\rb)}
{\alpha_{se}\alpha_{re}}-1-\frac{1}{\beta_{sd}}
+\frac{\alpha_{se}\alpha_{re}e^{-\beta_{sd}\lb(\rho-1\rb)}}{\beta_{sd}\lb(\alpha_{se}+\rho\beta_{sd}\rb)
\lb(\alpha_{re}+\rho\beta_{sd}\rb)}\rb].
\end{align}
}\\
{\centering The asymptotic SOP derived from (\ref{eq_20}) when $1/\beta_{rd}$ is fixed and
$1/\beta_{sr}=1/\beta\rightarrow  \infty$.
\begin{align}
\label{eq_55}
P_o^{AS}(R_s)&= \lb(1-\frac{\alpha_{se}\alpha_{re}}{\lb(\beta_{rd}-\beta_{sd}\rb)}\lb(\frac{\beta_{rd}e^{-\beta_{sd}(\rho-1)}}
{\lb(\alpha_{se}+\rho\beta_{sd}\rb)\lb(\alpha_{re}+\rho\beta_{sd}\rb)}
-\frac{\beta_{sd}e^{-\beta_{rd}(\rho-1)}}
{\lb(\alpha_{se}+\rho\beta_{rd}\rb)\lb(\alpha_{re}+\rho\beta_{rd}\rb)}\rb)\nn
\rb)\nn\\
&-\frac{\gamma_{th}}{\frac{1}{\beta}}
\lb[\frac{\alpha_{se}e^{-\beta_{sd}\lb(\rho-1\rb)}}{\alpha_{se}+\rho\beta_{sd}}-\frac{\alpha_{se}\alpha_{re}}{\lb(\beta_{rd}-\beta_{sd}\rb)}\lb(\frac{\beta_{rd}e^{-\beta_{sd}(\rho-1)}}
{\lb(\alpha_{se}+\rho\beta_{sd}\rb)\lb(\alpha_{re}+\rho\beta_{sd}\rb)}
-\frac{\beta_{sd}e^{-\beta_{rd}(\rho-1)}}
{\lb(\alpha_{se}+\rho\beta_{rd}\rb)\lb(\alpha_{re}+\rho\beta_{rd}\rb)}\rb)\rb].
\end{align}}
\\\hline
\end{tabular}
\end{table*}
\end{document}